\documentclass[aps,prl,twocolumn,showpacs]{revtex4}

\usepackage{graphicx}
\usepackage{dcolumn}
\usepackage{bm}
\usepackage{subfigure}
\usepackage{color}
\usepackage{hyperref}
\begin{document}

\title{Multiple Resonators as a Multi-Channel Bus for Coupling Josephson Junction Qubits }

\author{Z. E. Thrailkill}
\email{zet23@drexel.edu}
\author{J. G. Lambert}
\author{R. C. Ramos}

\affiliation{Department of Physics, Drexel University, Philadelphia, PA 19104 USA}
\homepage{http://www.physics.drexel.edu/research/lowtemp/}

\date{\today}

\begin{abstract}
Josephson junction-based qubits have been shown to be promising components for a future quantum computer.  A network of these superconducting qubits will require quantum information to be stored in and transferred among them.  Resonators made of superconducting metal strips are useful elements for this purpose because they have long coherence times and can dispersively couple qubits.  We explore the use of multiple resonators with different resonant frequencies to couple qubits.  We find that an array of resonators with different frequencies can be individually addressed to store and retrieve information, while coupling qubits dispersively. We show that a control qubit can be used to effectively isolate an active qubit from an array of resonators so that it can function within the same frequency range used by the resonators.   
\end{abstract}


\maketitle

There is currently a great deal of interest in building a working quantum computer due to its potentially powerful applications \cite{DavidDeutsch12081992,shor:1484,RevModPhys.74.145}.  One of the most promising approaches involves the use of superconducting qubits based on the Josephson junction \cite{devoret-2004}.  These solid-state systems have the advantage of being scaled up using well-established fabrication techniques.  However, many challenges remain in scaling up the number of qubits in any quantum computing system. These include the coupling of large numbers of qubits together and the storage of quantum information for later use.  Many theoretical and experimental studies have been done on the direct coupling of charge qubits \cite{Pashkin-2003,PhysRevB.68.024510}, phase qubits \cite{919517,R.McDermott02252005,steffen:050502,PhysRevLett.94.027003}, and flux qubits \cite{PhysRevLett.94.090501,PhysRevB.72.060506,niskanen:220503,grajcar:047006}.  The issues of how multiple qubits can be networked together and what entanglement protocols can correspondingly be implemented have also received a significant amount of attention \cite{galiautdinov:010305,strauch-2008-78,thrailkill-2009-3,thrailkill-JoP.150.052268}.  

	The use of resonators to hold quantum information and to couple qubits together is appealing due to their stability and long coherence times \cite{PhysRevB.68.064509}, which have been experimentally demonstrated \cite{wang:240401,citeulike:4697241,2008_Simmonds_TwoQubits,majer-2007-449}. This is expected because resonators are coupled via capacitors with no wires connected directly to the outside environment, making them more isolated than the qubit.   
	
	In this paper, we consider how an array of resonators can couple two Josephson junction qubits \cite{wallquist:224506}.  The resonators serve as a multi-channel bus to transfer information, as well as stable locations for the storage of quantum information.  However, we must look at the practical limitations of such a system.  First, we consider the frequency spacing of the resonator array, because the resonators will need to be sufficiently separated from each other to act as memory storage.  Second, we explore how the qubits can be dispersively coupled together via the resonator array.  Finally, we discuss selective coupling to individual resonators in the array.  This will allow the qubits to operate in the frequency band occupied by the array without coupling to any one resonator.  Here, we will focus on Josephson junction phase qubits. However, the results can be generalized in a straightforward way to flux- and charge-based qubits.

Any configuration of phase qubits coupled to resonators can be described by a generalized form of the Jaynes-Cummings Hamiltonian \cite{PhysRevA.69.062320}

\begin{eqnarray}
\label{Hamiltonian1}
H = \sum\limits_i {\frac{{\hbar \omega _{qi}}}{2}\hat \sigma _ + ^i } \hat \sigma _ - ^i  + \sum\limits_j {\hbar \omega _{rj}\hat a_j^\dag  \hat a_j } \nonumber\\+ \sum\limits_{i,j} {\hbar g_{ij} (\hat a_j^\dag  \hat \sigma _ - ^i  + \hat a_j \hat \sigma _ + ^i )} .
\end{eqnarray}

\noindent Here, the first and second terms are the energies of the qubits and resonators, respectively, and the third term describes the coupling between them.  The variable $\omega_{qi}$ is the frequency of qubit $i$ corresponding to the energy difference between its ground and excited states, $\omega_{rj}$ is the frequency of resonator $j$, and $g_{ij}$ is the coupling strength between qubit $i$ and resonator $j$.  The Pauli operators $\hat \sigma_+^i$ and $\hat \sigma_-^i$ add and remove excitations in qubit $i$, and  $\hat a_j^\dag$ and $\hat a_j$ are the creation and annihilation operators for resonator $j$, respectively.

Consider two qubits in resonance with each other and coupled through a resonator.  If the resonator is sufficiently detuned from the qubits, then dispersive coupling can occur \cite{blais:032329}.  In order to show this, we make a unitary transformation of the Hamiltonian: $UHU^\dag=H_{eff}$ where

\begin{equation}
U = \exp \left[ {\sum {\frac{{g_{ij} }}{{\Delta _{ij}
}}\left(\hat a_{j}^{\dag}  \hat \sigma _ - ^i  + \hat a_j \hat \sigma _ + ^i \right)}
} \right].
\end{equation}

Expanding the effective Hamiltonian to second order in the limit $ \left| {\Delta _{ij} } \right| = \left| {\omega {}_{qi} - \omega _{rj} } \right| \gg g_{ij} $ gives

 \begin{figure*}[htbp]
  	\begin{center}
		\subfigure[]{\label{block1}\includegraphics[scale=0.28]{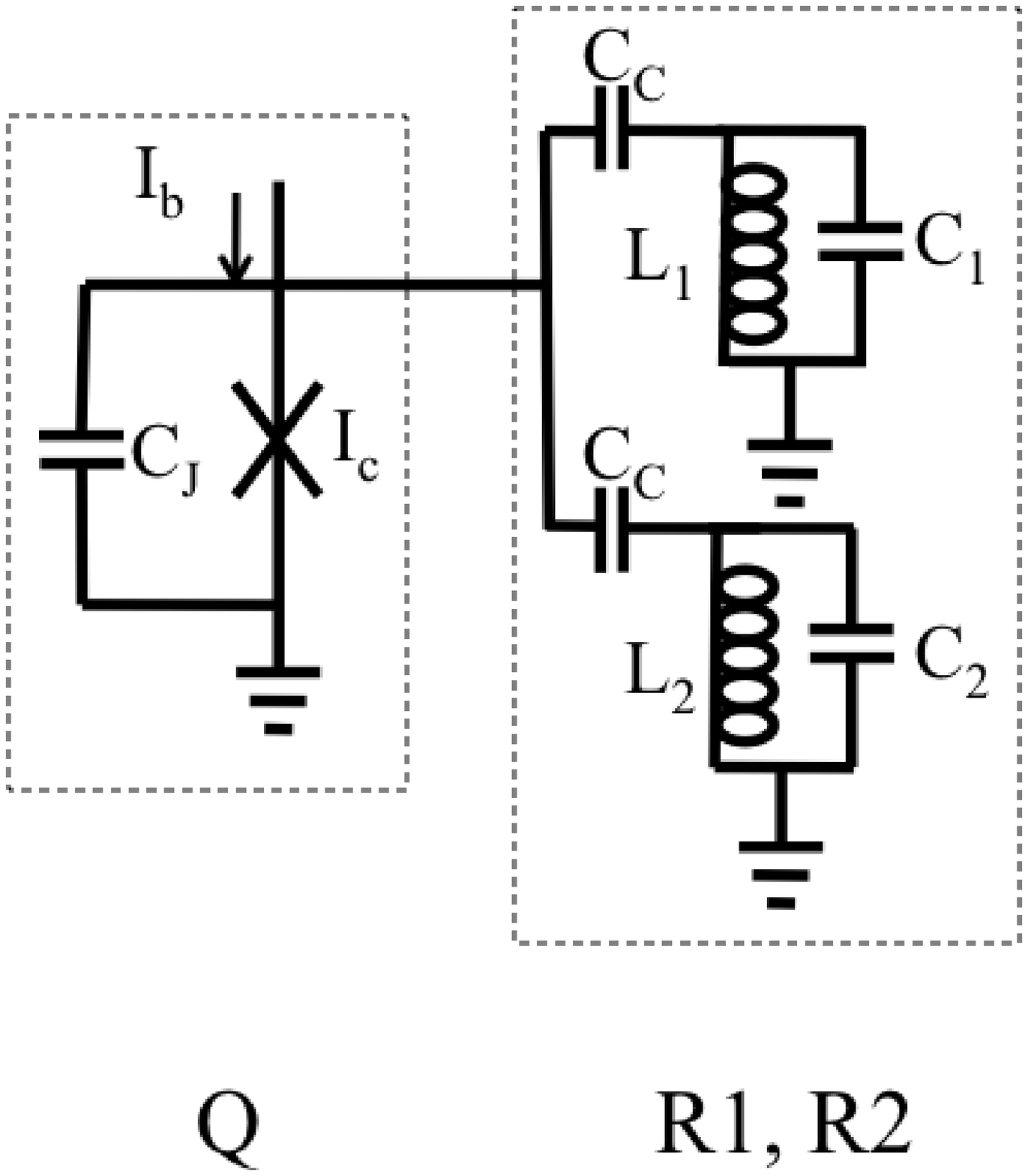}}
		\subfigure[]{\label{fig1}
		\includegraphics[width=2.8in]{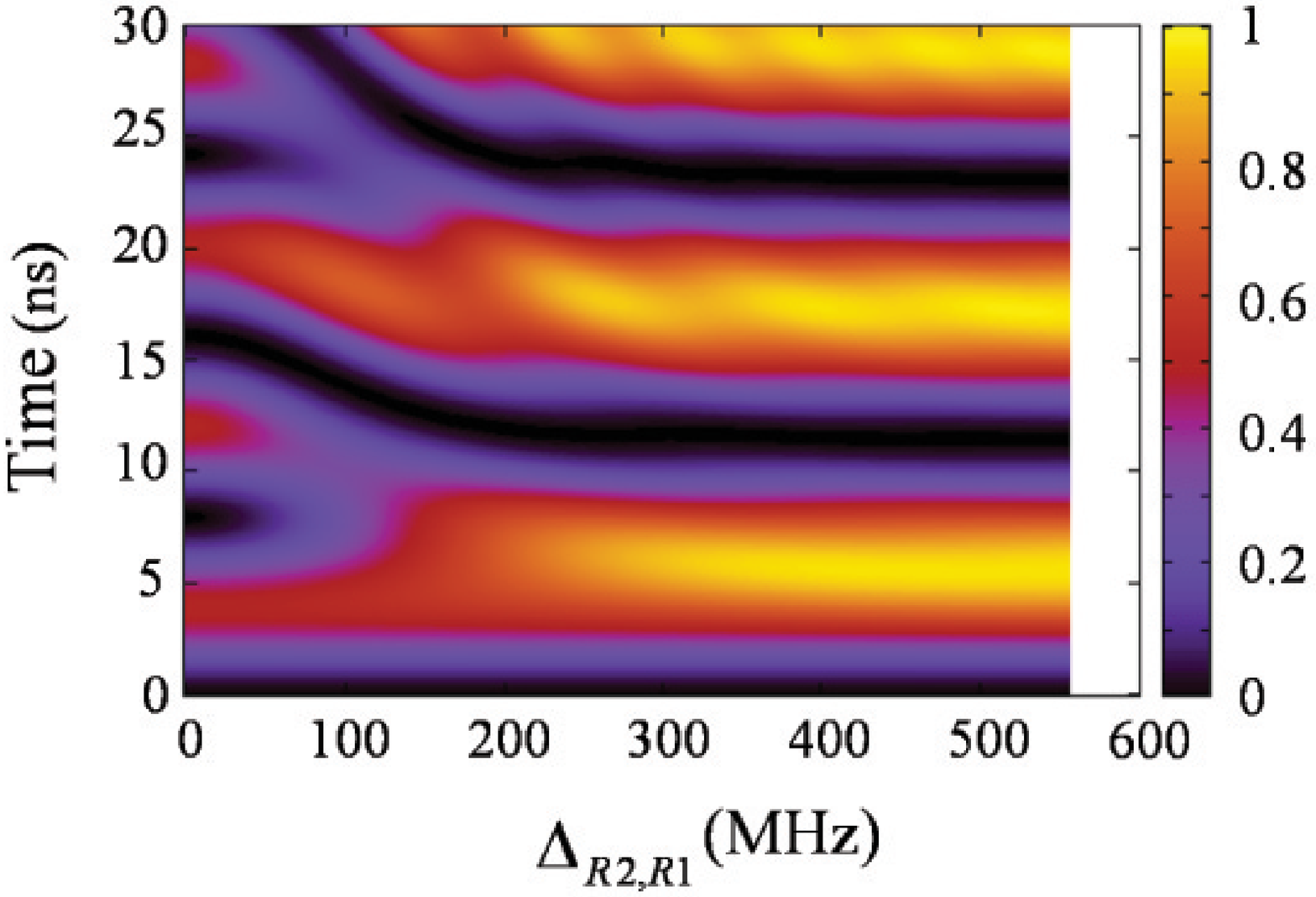}}
		
	\end{center}
\caption{ \label{figure1}(a) Schematic of a qubit (Q) coupled to an array of two resonators (R1, R2) via identical coupling capacitors $C_c$.  The qubit is characterized as having capacitance $C_J$ and critical current $I_c$, with bias current $I_b$.  The resonators are characterized by inductance $L_i$ and capacitance $C_i$, for $i=1,2$. 
(b) For this simulation, the system is initialized with a single excitation in Q, and Q is in-resonance with R1. Each vertical cut represents the population in R1 over time at a particular detuning $\Delta_{R2,R1}$, of R2 from R1.  The qubit-resonator coupling strength is 110 MHz.  At zero detuning, the excitation oscillates between the qubit and the two resonators; since the two resonators are identical here, each resonator is only half populated.  As R2 is detuned, the excitation oscillates between Q and R1 with minimal population in R2.  After more oscillations, R2 will accumulate some population, even at large detuning, which causes the appearance of ripples.}

\end{figure*}

\begin{widetext}
\begin{equation}
H_{eff}  = \sum\limits_i {\frac{{\hbar \omega_{qi}}}{2}\hat \sigma _ + ^i } \hat \sigma _ - ^i  + \sum\limits_j \left[{\hbar \left( {\omega_{rj} + \sum\limits_i {\frac{{g_{ij}^2 }}{{\Delta _{ij} }}\hat \sigma _ + ^i \hat \sigma _ - ^i } } \right)\hat a_j^\dag  \hat a_j }  + \sum\limits_{i > k} {\hbar \frac{{g_{ij} g_{kj} (\Delta _{ij}  + \Delta _{kj} )}}{{2\Delta _{ij} \Delta _{kj} }}(\hat \sigma _ - ^i \hat \sigma _ + ^k  + \hat \sigma _ + ^i \hat \sigma _ - ^k )} \right].
\end{equation}
\end{widetext}

\noindent Here, the first term contains the individual qubit energies, while the second term contains the resonator energies.  It is interesting to note that the resonator term contains the qubit operators.  This term translates into a dispersive, qubit state-dependent shift in the resonator frequencies  \cite{koch-2007-76} \footnote{This is a useful feature for the non-demolition readout of the qubit state.}.  The last term represents the dispersive coupling between the qubits.  We perform simulations, as described in what follows, by implementing the Hamiltonian in Eq. (\ref{Hamiltonian1}), using the procedures outlined in Ref. \cite{thrailkill-2009-3}.

We consider an array of resonators used as a memory register.  To accomplish this, one must be able to transfer an excitation from a qubit to a specific resonator, without coupling to the other resonators in the array.  This is implemented by designing resonators that are sufficiently detuned from each other.  To determine the amount of detuning required to avoid crosstalk between resonators, we examine a single qubit coupled to an array of two resonators as shown in Fig. \ref{block1}.  The qubit (Q) and resonator 1 (R1) are fixed at the same frequency while resonator 2 (R2) is detuned.  All qubit-resonator couplings are presumed to be identical with a strength of $g_{ij}=110 \text{ MHz}$ for all $i,j$, i.e., all coupling capacitances $C_c$ are equal.  This value is typical in experiments such as in Ref. \cite{majer-2007-449}.  We begin with Q in the excited state and let the system evolve over time while recording the population in R1, as we increase the detuning $\Delta_{R2,R1}$, of R2.  The result of this simulation is shown in Fig. \ref{fig1}.    

At zero detuning,  the population in R1 oscillates between approximately 0 and 0.5.  This is because the system is oscillating between states $\left|010\right>$ and $\left(\left|100\right> + \left|001\right>\right)/\sqrt{2}$ where $\left|lmn\right>=\left|l\right>_{R 1}\bigotimes \left|m\right>_{Q}\bigotimes \left|n\right>_{R 2} $.  When $\Delta_{R2,R1}\approx 400$ MHz and higher, all the excitation is transferred into R1 in about $t = 5 \text{ ns}$.  This transfer time depends on the qubit-resonator coupling strength, which can be chosen based on the desired time scale, e.g., the amount of time necessary to complete gate operations.  The required detuning $\Delta_{R2,R1}$ is dependent on the qubit-resonator coupling strength $g_{ij}$.  In this case, with $\Delta_{R2,R1}\approx 400$ MHz and $g_{ij}=110$ MHz, the required detuning needs to be about a factor of four larger than the qubit-resonator coupling strength.  We have performed similar studies of systems with up to 6 resonators, with equal detuning between consecutive resonators, and identical qubit-resonator coupling strengths.  The minimum required detuning for any given time scale increases with the number of resonators, indicating that the relative detuning is in fact system dependent.  As a result, the minimum detuning for resonator isolation must be carefully analyzed for each system.

\begin{figure*}[htbp]
\centering
	\begin{minipage}[b]{3.1 in}
			\subfigure[]{\label{block2}\includegraphics[scale=0.3]{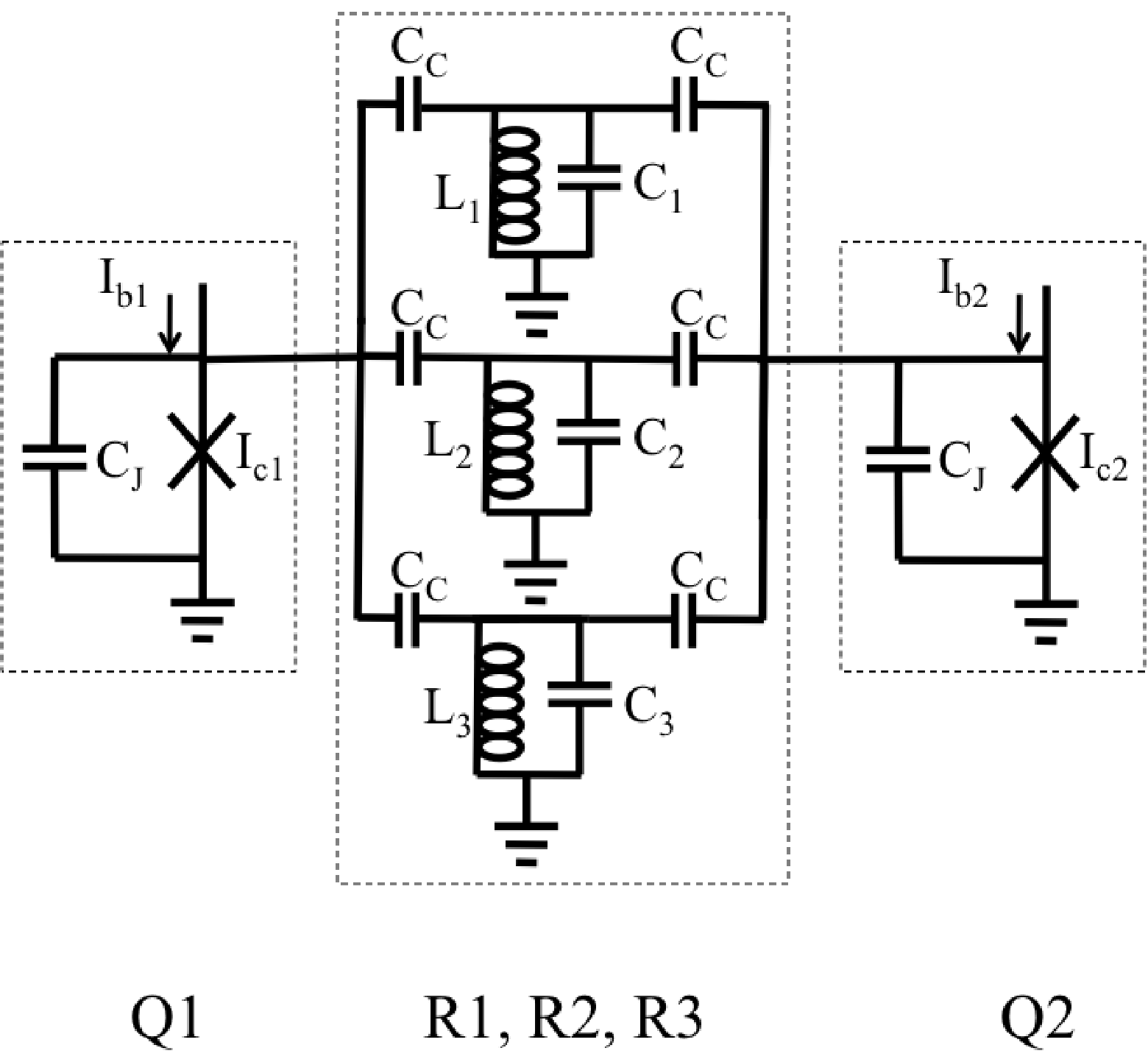}}
			\vspace{.8 in}
	\end{minipage}
	\hspace{.5cm}
	\begin{minipage}[b]{3.1 in}
			
			\subfigure[]{\label{fig2}
			\includegraphics[width=3in]{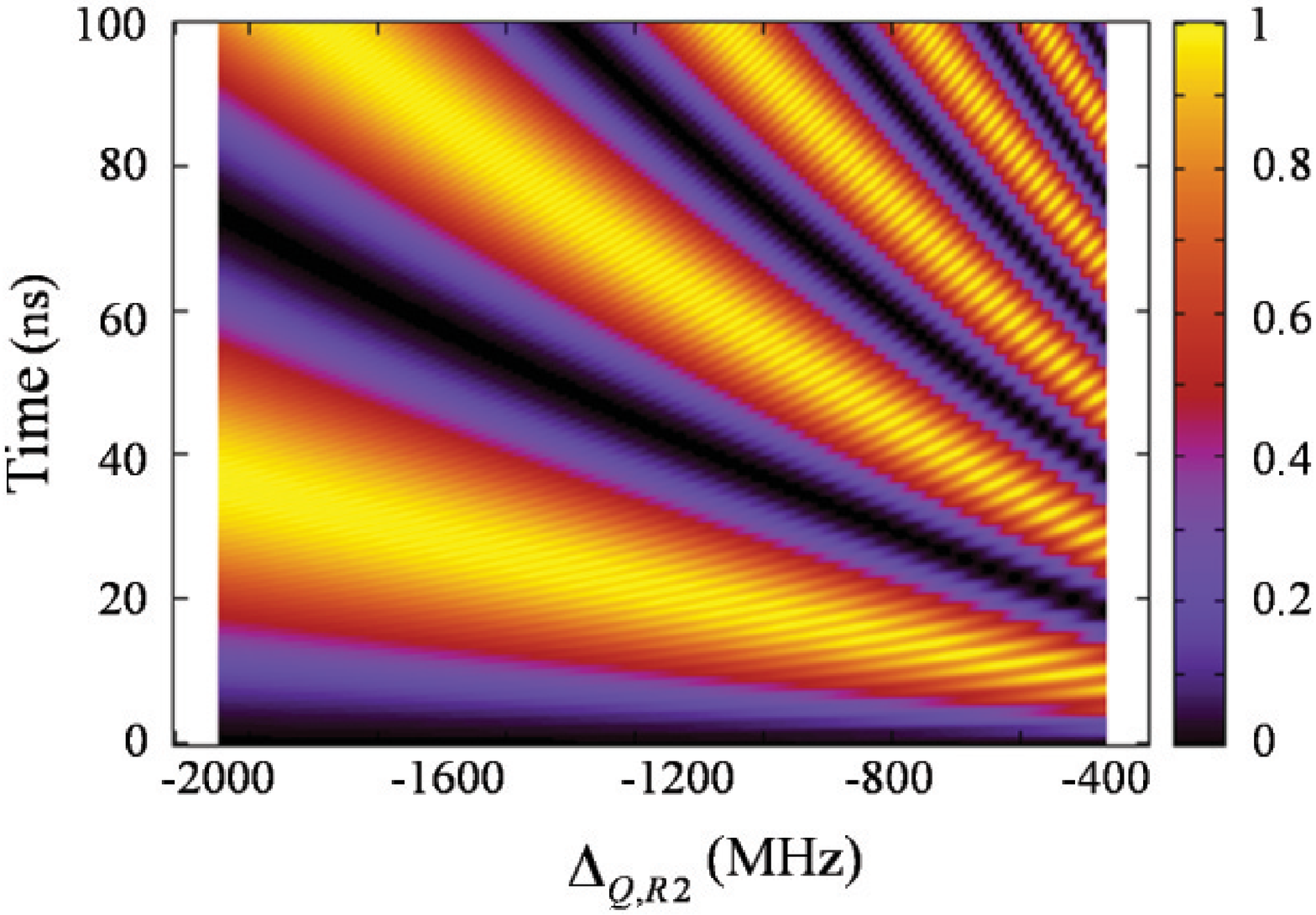}}
			\\
			\vspace{-0.1 in}
			\subfigure[]{\label{fig3}\includegraphics[width=3in]{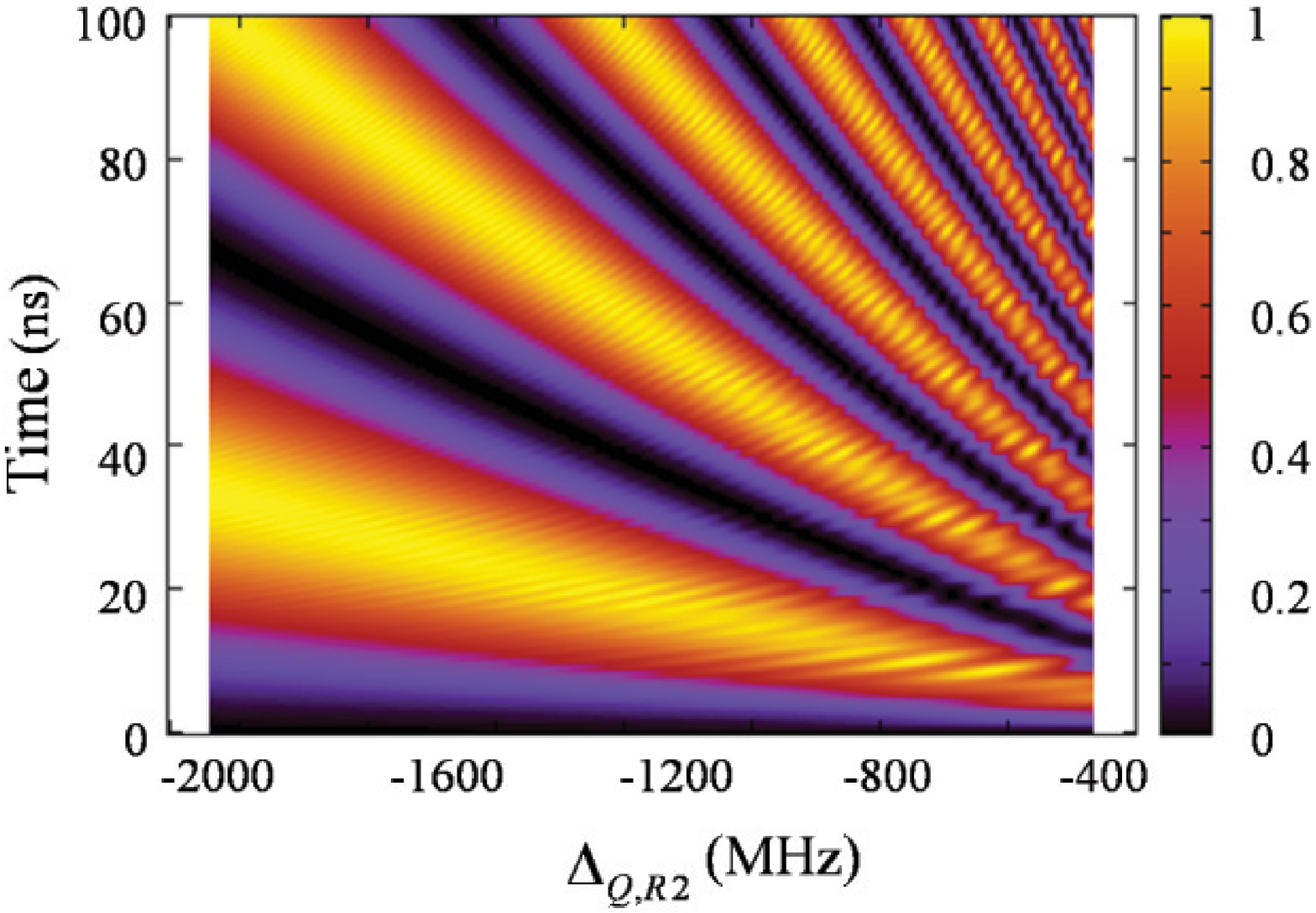}}
	\end{minipage}
	\vspace{-0.1 in}
\caption{\label{figure2}(a) Two qubits (Q1, Q2) dispersively coupled to an array of three resonators (R1, R2, R3) via identical coupling capacitors $C_c$.  (b),(c)  For these two simulations, the system is initialized with a single excitation in Q1, and the two qubits maintain equal frequencies as they are simultaneously detuned from the resonators.  Each vertical cut represents the population in R2 over time at a particular detuning $\Delta_{Q,R2}$, of the qubits from R2.  In (b), the frequencies of all three resonators are equal ($\omega_{R1}=\omega_{R2}=\omega_{R3}$).  At large detuning ($\Delta_{Q,R2}=$ -2000 MHz), the excitation smoothly oscillates between the two qubits without significant interference from the resonators.  As the magnitude of the detuning decreases, the effective coupling between the two qubits strengthens, thus the oscillation of the excitation becomes more frequent.  Also, the direct coupling of the qubits to the resonators strengthens, causing the small ripples.  In (c), the frequencies of R1 and R3 are set  slightly above and below R2, respectively ($\omega_{R1}>\omega_{R2}>\omega_{R3}$).  The excitation oscillates slightly faster than in (b) because the small offsets in frequency of R1 and R3 increases the coupling bandwidth, resulting in a small increase in coupling between the qubits over the same range of detuning.  The offset of R1 and R3 from R2 also causes the ripples to be non-uniform at smaller detuning.}
\end{figure*}

Next, we investigate the behavior of a system consisting of two qubits dispersively coupled to an array of three resonators, as shown in Fig. \ref{block2}.  We consider two cases: (1) All three resonators are designed with the same resonant frequency, and (2) the resonant frequencies of the three resonators are slightly detuned so that $\omega_{R1}>\omega_{R2}>\omega_{R3}$.   We demonstrate how information in the form of an excitation is transferred dispersively from one qubit to another through an array of resonators.  In both cases, the system is initialized with a single excitation in Q1, and both qubits are held in resonance with each other as the magnitude of their detuning from R2 is decreased from, say, -2000 MHz to -400 MHz.

In both cases, at large detuning of -2000 MHz, the excitation appears to smoothly oscillate into qubit 2, as shown in Fig. \ref{fig2} and \ref{fig3}.  As the detuning decreases, two things happen:  First, the time it takes for the excitation to move into qubit 2 gets shorter, indicating an increase in the effective coupling strength between the two qubits.  Second, small ripples, or oscillations, begin to appear.  These oscillations are due to the detuning becoming small enough that the dispersive limit approximation starts to break down.  This result shows us that using three identical resonators in parallel to dispersively couple qubits is effectively similar to using only one resonator with three times the coupling strength.  In general, an array of $n$ resonators can be replaced by a single resonator with $n$ times the coupling strength.  The dispersive coupling can be increased by reducing the detuning, but at the cost of having significant oscillations between the qubits and resonators.  

\begin{figure*}[htbp]
	\begin{center}
		\subfigure[]{\label{block3}\includegraphics[scale=0.28]{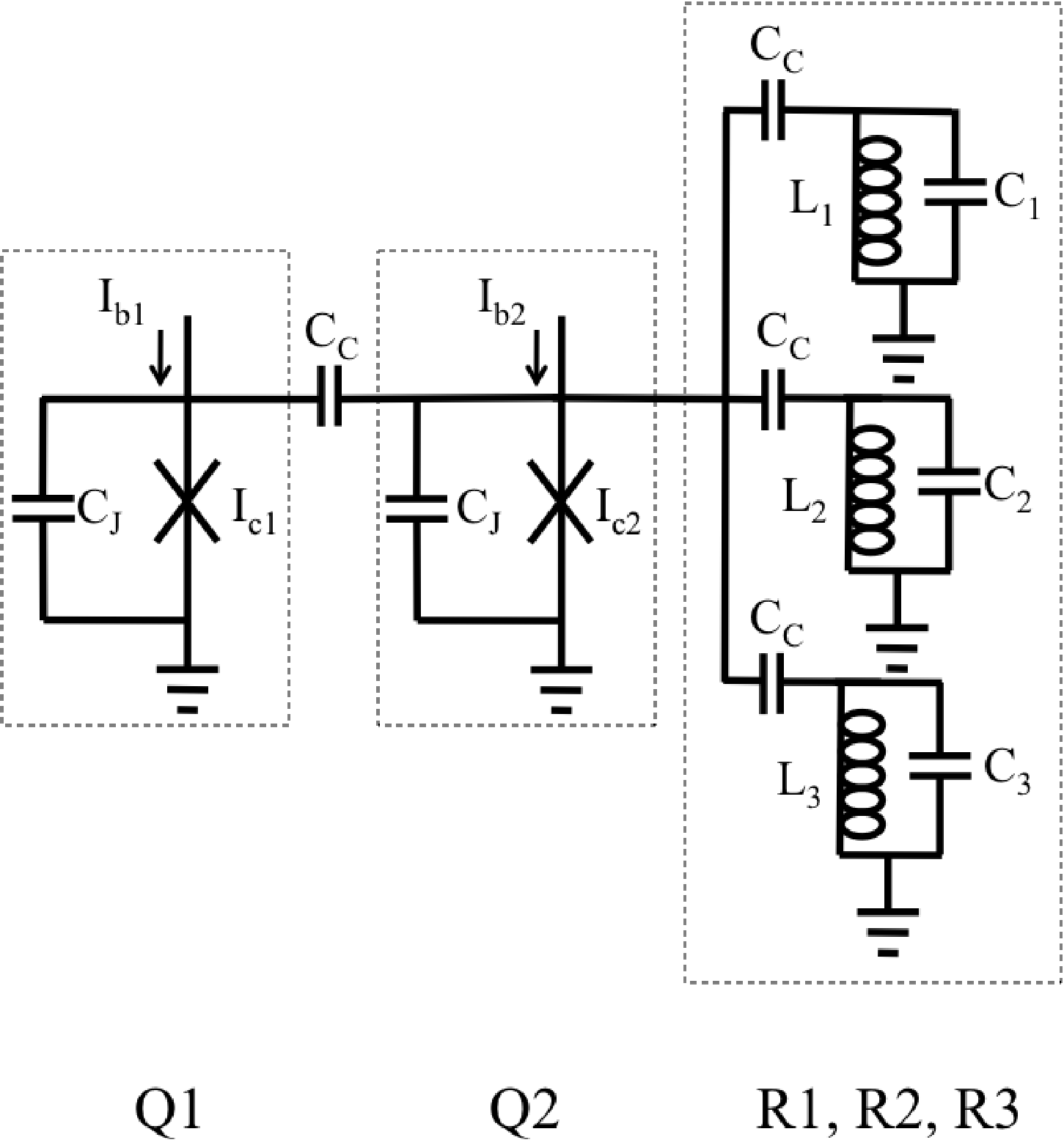}}
		\subfigure[]{\label{fig4}\includegraphics[width=2.8in]{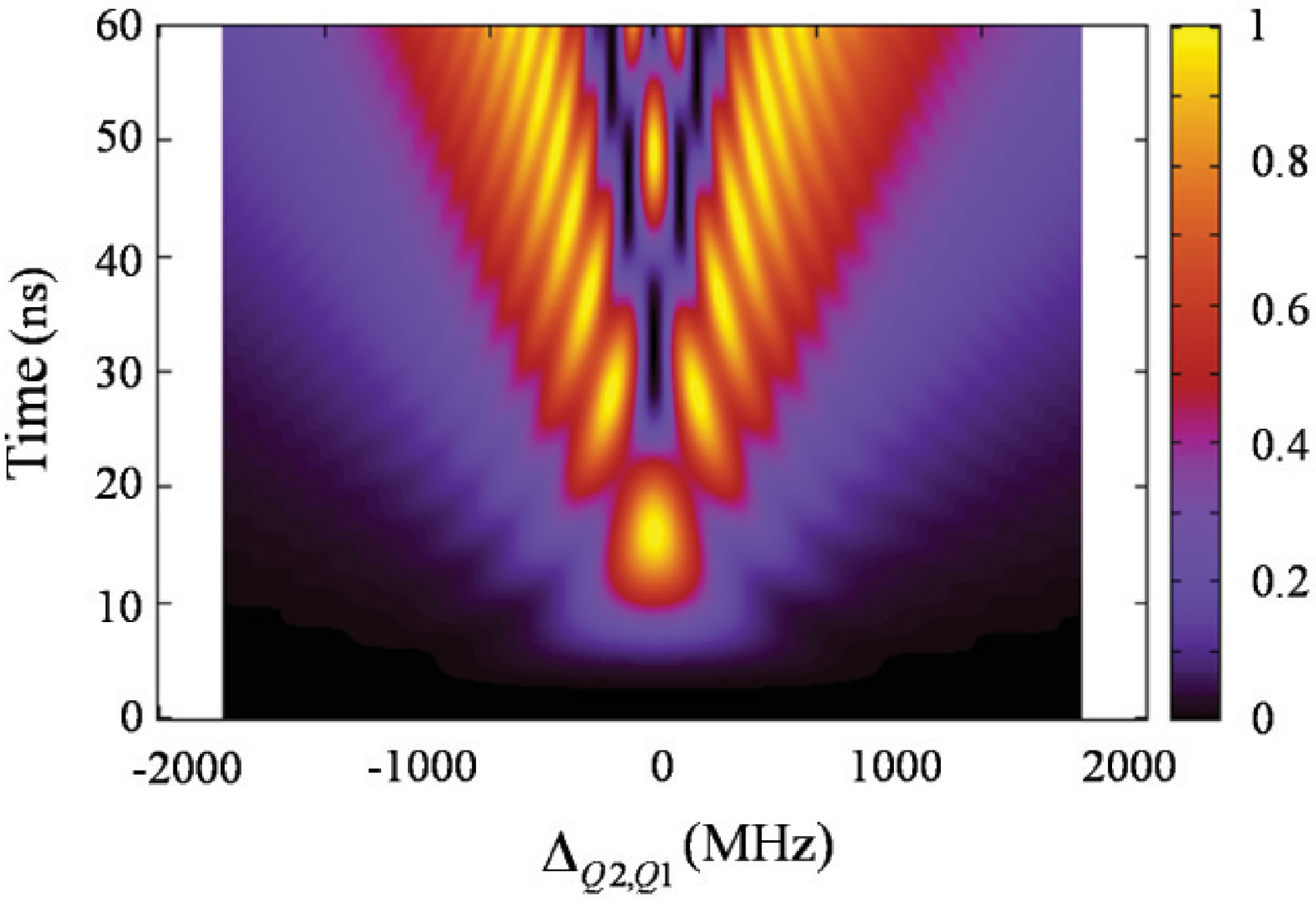}}
	\end{center}
\caption{\label{figure3} (a)  Placing a control qubit (Q2) between the active qubit (Q1) and the resonator array (R1, R2, R3) allows coupling to be turned on and off.  (b) The system is initialized with an excitation in Q1, Q1 is in-resonance with R2, and the frequencies of R1 and R3 are set slightly above and below R2, respectively ($\omega_{R1}>\omega_{R2}>\omega_{R3}$).  Each vertical cut represents the population in R2 over time at a particular detuning $\Delta_{Q2,Q1}$, of Q2 from Q1.  At zero detuning, the excitation readily oscillates in and out of the resonator R2.  As Q2 is detuned further away, the coupling between Q1 and R2 becomes weaker, resulting in slower oscillations of the excitation.  The detuning of the control qubit Q2 can be chosen based on the desired time scale, e.g. the time required to manipulate Q1.}
\end{figure*}

The second case where $\omega_{R1}>\omega_{R2}>\omega_{R3}$  is shown in Fig. \ref{fig3}.   This is very similar to the first case, however, there is a slight  overall increase in the state transfer rate.  This is because the detuning is still relative to R2, but the detuning of R3 is now slightly less.  As the qubits approach resonance with the resonators from below, the dispersive coupling strength becomes slightly larger because the frequency of R3 is a little closer to the frequency of the qubits.  The small differences in resonator frequencies also cause slight non-uniformities in the high frequency oscillations or ripples that become more pronounced at lower detuning.  This interference is not a factor if the qubits are sufficiently detuned. Regardless of these artifacts, it is clear that an array of resonators that is used for memory storage can also be used to dispersively couple qubits.  

The state transfer between two superconducting qubits via a single resonator in the non-dispersive regime has  been demonstrated in elegant experiments  by Sillanp\"{a}\"{a} \emph{et al}. \cite{2008_Simmonds_TwoQubits}.  However, this method requires the excitation to be completely transferred into the resonator.  State transfer via dispersive coupling, as demonstrated by Majer \emph{et al}. \cite{majer-2007-449}, does not require significant energy transfer to the resonators.  This allows the resonators to be used to couple qubits and to store state information, concurrently.

In systems with a large array of resonators, there is a considerably large bandwidth used up, within which qubits cannot be operated without directly coupling to the resonators.  This problem can be solved by introducing a ``control'' qubit placed between the resonator array and the active qubit.  This configuration is shown in Fig. \ref{block3} where Q1 is the active qubit and Q2 is the control qubit.  In the simulation, the three resonators are identical to the ones in the previous discussion where $\omega_{R1}>\omega_{R2}>\omega_{R3}$.  Q1 is held in resonance with R2 and Q1 is initialize in its excited state.  We can enable or disable transmission of the excitation to R2 by tuning Q2.

The result of the simulation is shown in Fig. \ref{fig4} where the population in R2 is plotted versus the detuning $\Delta_{Q2,Q1}$, of Q2 from Q1, in the range of -2000 MHz to 2000 MHz.  As per design, at zero detuning, the excitation is transferred after a time of 16 ns. This time can be chosen based on the desired time scale by adjusting the qubit-qubit and qubit-resonator coupling strengths.  At large detuning, i.e., beyond $\pm1.5$ GHz, the coupling between Q1 and R2 becomes dispersive up to about 30 ns, showing that dispersive coupling can be weak enough to isolate the active qubit.  Q2 can be detuned further to reduce the dispersive coupling if the desired time scale is longer, e.g., to perform operations on Q1.  Thus, this simulation shows that a control qubit can be effectively used to turn coupling on and off between a qubit and an array of resonators.

In conclusion, we have examined how an array of resonators can be utilized as both a memory register and as a multi-channel bus for quantum information transfer.  We have found that the resonators must be properly spaced apart in frequency to remain effectively isolated from one another.  This means that there will be a significant bandwidth of frequencies that is occupied by the resonators.  We have found that, despite this, the resonator array can still be successfully used to dispersively couple qubits at large detuning values.  Furthermore, to employ the resonators for memory, a control qubit can be used to selectively couple to individual resonators, thus gaining the benefit of a resonator memory bank, while allowing the qubits to operate in the same frequency range.

\bibliography{ref}

\begin{thebibliography}{27}
\expandafter\ifx\csname natexlab\endcsname\relax\def\natexlab#1{#1}\fi
\expandafter\ifx\csname bibnamefont\endcsname\relax
  \def\bibnamefont#1{#1}\fi
\expandafter\ifx\csname bibfnamefont\endcsname\relax
  \def\bibfnamefont#1{#1}\fi
\expandafter\ifx\csname citenamefont\endcsname\relax
  \def\citenamefont#1{#1}\fi
\expandafter\ifx\csname url\endcsname\relax
  \def\url#1{\texttt{#1}}\fi
\expandafter\ifx\csname urlprefix\endcsname\relax\def\urlprefix{URL }\fi
\providecommand{\bibinfo}[2]{#2}
\providecommand{\eprint}[2][]{\url{#2}}

\bibitem[{\citenamefont{Deutsch and Jozsa}(1992)}]{DavidDeutsch12081992}
\bibinfo{author}{\bibfnamefont{D.}~\bibnamefont{Deutsch}} \bibnamefont{and}
  \bibinfo{author}{\bibfnamefont{R.}~\bibnamefont{Jozsa}},
  \bibinfo{journal}{Proc. R. Soc. London A} \textbf{\bibinfo{volume}{439}},
  \bibinfo{pages}{553} (\bibinfo{year}{1992}).

\bibitem[{\citenamefont{Shor}(1997)}]{shor:1484}
\bibinfo{author}{\bibfnamefont{P.~W.} \bibnamefont{Shor}},
  \bibinfo{journal}{SIAM Journal on Computing} \textbf{\bibinfo{volume}{26}},
  \bibinfo{pages}{1484} (\bibinfo{year}{1997}).

\bibitem[{\citenamefont{Gisin et~al.}(2002)\citenamefont{Gisin, Ribordy,
  Tittel, and Zbinden}}]{RevModPhys.74.145}
\bibinfo{author}{\bibfnamefont{N.}~\bibnamefont{Gisin}},
  \bibinfo{author}{\bibfnamefont{G.}~\bibnamefont{Ribordy}},
  \bibinfo{author}{\bibfnamefont{W.}~\bibnamefont{Tittel}}, \bibnamefont{and}
  \bibinfo{author}{\bibfnamefont{H.}~\bibnamefont{Zbinden}},
  \bibinfo{journal}{Rev. Mod. Phys.} \textbf{\bibinfo{volume}{74}},
  \bibinfo{pages}{145} (\bibinfo{year}{2002}).

\bibitem[{\citenamefont{{Devoret} et~al.}(2004)\citenamefont{{Devoret},
  {Wallraff}, and {Martinis}}}]{devoret-2004}
\bibinfo{author}{\bibfnamefont{M.~H.} \bibnamefont{{Devoret}}},
  \bibinfo{author}{\bibfnamefont{A.}~\bibnamefont{{Wallraff}}},
  \bibnamefont{and} \bibinfo{author}{\bibfnamefont{J.~M.}
  \bibnamefont{{Martinis}}}, \bibinfo{journal}{ArXiv e-prints}
  (\bibinfo{year}{2004}), \eprint{arXiv:cond-mat/0411174}.

\bibitem[{\citenamefont{Pashkin et~al.}(2003)\citenamefont{Pashkin, Yamamoto,
  Astafiev, Nakamura, Averin, and Tsai}}]{Pashkin-2003}
\bibinfo{author}{\bibfnamefont{Y.~A.} \bibnamefont{Pashkin}},
  \bibinfo{author}{\bibfnamefont{T.}~\bibnamefont{Yamamoto}},
  \bibinfo{author}{\bibfnamefont{O.}~\bibnamefont{Astafiev}},
  \bibinfo{author}{\bibfnamefont{Y.}~\bibnamefont{Nakamura}},
  \bibinfo{author}{\bibfnamefont{D.~V.} \bibnamefont{Averin}},
  \bibnamefont{and} \bibinfo{author}{\bibfnamefont{J.~S.} \bibnamefont{Tsai}},
  \bibinfo{journal}{Nature} \textbf{\bibinfo{volume}{421}},
  \bibinfo{pages}{823} (\bibinfo{year}{2003}).

\bibitem[{\citenamefont{You et~al.}(2003)\citenamefont{You, Tsai, and
  Nori}}]{PhysRevB.68.024510}
\bibinfo{author}{\bibfnamefont{J.~Q.} \bibnamefont{You}},
  \bibinfo{author}{\bibfnamefont{J.~S.} \bibnamefont{Tsai}}, \bibnamefont{and}
  \bibinfo{author}{\bibfnamefont{F.}~\bibnamefont{Nori}},
  \bibinfo{journal}{Phys. Rev. B} \textbf{\bibinfo{volume}{68}},
  \bibinfo{pages}{024510} (\bibinfo{year}{2003}).

\bibitem[{\citenamefont{Ramos et~al.}(2001)\citenamefont{Ramos, Gubrud,
  Berkley, Anderson, Lobb, and Wellstood}}]{919517}
\bibinfo{author}{\bibfnamefont{R.~C.} \bibnamefont{Ramos}},
  \bibinfo{author}{\bibfnamefont{M.~A.} \bibnamefont{Gubrud}},
  \bibinfo{author}{\bibfnamefont{A.~J.} \bibnamefont{Berkley}},
  \bibinfo{author}{\bibfnamefont{J.~R.} \bibnamefont{Anderson}},
  \bibinfo{author}{\bibfnamefont{C.~J.} \bibnamefont{Lobb}}, \bibnamefont{and}
  \bibinfo{author}{\bibfnamefont{F.~C.} \bibnamefont{Wellstood}},
  \bibinfo{journal}{IEEE Trans. Appl. Supercond.}
  \textbf{\bibinfo{volume}{11}}, \bibinfo{pages}{998} (\bibinfo{year}{2001}).

\bibitem[{\citenamefont{McDermott et~al.}(2005)\citenamefont{McDermott,
  Simmonds, Steffen, Cooper, Cicak, Osborn, Oh, Pappas, and
  Martinis}}]{R.McDermott02252005}
\bibinfo{author}{\bibfnamefont{R.}~\bibnamefont{McDermott}},
  \bibinfo{author}{\bibfnamefont{R.~W.} \bibnamefont{Simmonds}},
  \bibinfo{author}{\bibfnamefont{M.}~\bibnamefont{Steffen}},
  \bibinfo{author}{\bibfnamefont{K.~B.} \bibnamefont{Cooper}},
  \bibinfo{author}{\bibfnamefont{K.}~\bibnamefont{Cicak}},
  \bibinfo{author}{\bibfnamefont{K.~D.} \bibnamefont{Osborn}},
  \bibinfo{author}{\bibfnamefont{S.}~\bibnamefont{Oh}},
  \bibinfo{author}{\bibfnamefont{D.~P.} \bibnamefont{Pappas}},
  \bibnamefont{and} \bibinfo{author}{\bibfnamefont{J.~M.}
  \bibnamefont{Martinis}}, \bibinfo{journal}{Science}
  \textbf{\bibinfo{volume}{307}}, \bibinfo{pages}{1299} (\bibinfo{year}{2005}).

\bibitem[{\citenamefont{Steffen et~al.}(2006)\citenamefont{Steffen, Ansmann,
  McDermott, Katz, Bialczak, Lucero, Neeley, Weig, Cleland, and
  Martinis}}]{steffen:050502}
\bibinfo{author}{\bibfnamefont{M.}~\bibnamefont{Steffen}},
  \bibinfo{author}{\bibfnamefont{M.}~\bibnamefont{Ansmann}},
  \bibinfo{author}{\bibfnamefont{R.}~\bibnamefont{McDermott}},
  \bibinfo{author}{\bibfnamefont{N.}~\bibnamefont{Katz}},
  \bibinfo{author}{\bibfnamefont{R.~C.} \bibnamefont{Bialczak}},
  \bibinfo{author}{\bibfnamefont{E.}~\bibnamefont{Lucero}},
  \bibinfo{author}{\bibfnamefont{M.}~\bibnamefont{Neeley}},
  \bibinfo{author}{\bibfnamefont{E.~M.} \bibnamefont{Weig}},
  \bibinfo{author}{\bibfnamefont{A.~N.} \bibnamefont{Cleland}},
  \bibnamefont{and} \bibinfo{author}{\bibfnamefont{J.~M.}
  \bibnamefont{Martinis}}, \bibinfo{journal}{Phys. Rev. Lett.}
  \textbf{\bibinfo{volume}{97}}, \bibinfo{eid}{050502} (\bibinfo{year}{2006}).

\bibitem[{\citenamefont{Xu et~al.}(2005)}]{PhysRevLett.94.027003}
\bibinfo{author}{\bibfnamefont{H.}~\bibnamefont{Xu}} \bibnamefont{et~al.},
  \bibinfo{journal}{Phys. Rev. Lett.} \textbf{\bibinfo{volume}{94}},
  \bibinfo{pages}{027003} (\bibinfo{year}{2005}).

\bibitem[{\citenamefont{Majer et~al.}(2005)\citenamefont{Majer, Paauw, ter
  Haar, Harmans, and Mooij}}]{PhysRevLett.94.090501}
\bibinfo{author}{\bibfnamefont{J.~B.} \bibnamefont{Majer}},
  \bibinfo{author}{\bibfnamefont{F.~G.} \bibnamefont{Paauw}},
  \bibinfo{author}{\bibfnamefont{A.~C.~J.} \bibnamefont{ter Haar}},
  \bibinfo{author}{\bibfnamefont{C.~J. P.~M.} \bibnamefont{Harmans}},
  \bibnamefont{and} \bibinfo{author}{\bibfnamefont{J.~E.} \bibnamefont{Mooij}},
  \bibinfo{journal}{Phys. Rev. Lett.} \textbf{\bibinfo{volume}{94}},
  \bibinfo{pages}{090501} (\bibinfo{year}{2005}).

\bibitem[{\citenamefont{Plourde et~al.}(2005)\citenamefont{Plourde, Robertson,
  Reichardt, Hime, Linzen, Wu, and Clarke}}]{PhysRevB.72.060506}
\bibinfo{author}{\bibfnamefont{B.~L.~T.} \bibnamefont{Plourde}},
  \bibinfo{author}{\bibfnamefont{T.~L.} \bibnamefont{Robertson}},
  \bibinfo{author}{\bibfnamefont{P.~A.} \bibnamefont{Reichardt}},
  \bibinfo{author}{\bibfnamefont{T.}~\bibnamefont{Hime}},
  \bibinfo{author}{\bibfnamefont{S.}~\bibnamefont{Linzen}},
  \bibinfo{author}{\bibfnamefont{C.-E.} \bibnamefont{Wu}}, \bibnamefont{and}
  \bibinfo{author}{\bibfnamefont{J.}~\bibnamefont{Clarke}},
  \bibinfo{journal}{Phys. Rev. B} \textbf{\bibinfo{volume}{72}},
  \bibinfo{pages}{060506} (\bibinfo{year}{2005}).

\bibitem[{\citenamefont{Niskanen et~al.}(2006)\citenamefont{Niskanen, Harrabi,
  Yoshihara, Nakamura, and Tsai}}]{niskanen:220503}
\bibinfo{author}{\bibfnamefont{A.~O.} \bibnamefont{Niskanen}},
  \bibinfo{author}{\bibfnamefont{K.}~\bibnamefont{Harrabi}},
  \bibinfo{author}{\bibfnamefont{F.}~\bibnamefont{Yoshihara}},
  \bibinfo{author}{\bibfnamefont{Y.}~\bibnamefont{Nakamura}}, \bibnamefont{and}
  \bibinfo{author}{\bibfnamefont{J.~S.} \bibnamefont{Tsai}},
  \bibinfo{journal}{Phys. Rev. B} \textbf{\bibinfo{volume}{74}},
  \bibinfo{eid}{220503} (\bibinfo{year}{2006}).

\bibitem[{\citenamefont{Grajcar et~al.}(2006)}]{grajcar:047006}
\bibinfo{author}{\bibfnamefont{M.}~\bibnamefont{Grajcar}} \bibnamefont{et~al.},
  \bibinfo{journal}{Phys. Rev. Lett.} \textbf{\bibinfo{volume}{96}},
  \bibinfo{eid}{047006} (\bibinfo{year}{2006}).

\bibitem[{\citenamefont{Galiautdinov and Martinis}(2008)}]{galiautdinov:010305}
\bibinfo{author}{\bibfnamefont{A.}~\bibnamefont{Galiautdinov}}
  \bibnamefont{and} \bibinfo{author}{\bibfnamefont{J.~M.}
  \bibnamefont{Martinis}}, \bibinfo{journal}{Phys. Rev. A}
  \textbf{\bibinfo{volume}{78}}, \bibinfo{eid}{010305} (\bibinfo{year}{2008}).

\bibitem[{\citenamefont{Strauch and Williams}(2008)}]{strauch-2008-78}
\bibinfo{author}{\bibfnamefont{F.~W.} \bibnamefont{Strauch}} \bibnamefont{and}
  \bibinfo{author}{\bibfnamefont{C.~J.} \bibnamefont{Williams}},
  \bibinfo{journal}{Phys. Rev. B} \textbf{\bibinfo{volume}{78}},
  \bibinfo{pages}{094516} (\bibinfo{year}{2008}).

\bibitem[{\citenamefont{Thrailkill
  et~al.}(2009{\natexlab{a}})\citenamefont{Thrailkill, Kennerly, and
  Ramos}}]{thrailkill-2009-3}
\bibinfo{author}{\bibfnamefont{Z.~E.} \bibnamefont{Thrailkill}},
  \bibinfo{author}{\bibfnamefont{S.~T.} \bibnamefont{Kennerly}},
  \bibnamefont{and} \bibinfo{author}{\bibfnamefont{R.~C.} \bibnamefont{Ramos}},
  \bibinfo{journal}{IEEE Trans. Appl. Supercond.}
  \textbf{\bibinfo{volume}{19}}, \bibinfo{pages}{968}
  (\bibinfo{year}{2009}{\natexlab{a}}).

\bibitem[{\citenamefont{Thrailkill
  et~al.}(2009{\natexlab{b}})\citenamefont{Thrailkill, Kennerly, Tyler, and
  Ramos}}]{thrailkill-JoP.150.052268}
\bibinfo{author}{\bibfnamefont{Z.}~\bibnamefont{Thrailkill}},
  \bibinfo{author}{\bibfnamefont{S.}~\bibnamefont{Kennerly}},
  \bibinfo{author}{\bibfnamefont{A.}~\bibnamefont{Tyler}}, \bibnamefont{and}
  \bibinfo{author}{\bibfnamefont{R.~C.} \bibnamefont{Ramos}},
  \bibinfo{journal}{J. Phys. Conf. Ser.} \textbf{\bibinfo{volume}{150}},
  \bibinfo{pages}{052268} (\bibinfo{year}{2009}{\natexlab{b}}).

\bibitem[{\citenamefont{You and Nori}(2003)}]{PhysRevB.68.064509}
\bibinfo{author}{\bibfnamefont{J.~Q.} \bibnamefont{You}} \bibnamefont{and}
  \bibinfo{author}{\bibfnamefont{F.}~\bibnamefont{Nori}},
  \bibinfo{journal}{Phys. Rev. B} \textbf{\bibinfo{volume}{68}},
  \bibinfo{pages}{064509} (\bibinfo{year}{2003}).

\bibitem[{\citenamefont{Wang et~al.}(2008)}]{wang:240401}
\bibinfo{author}{\bibfnamefont{H.}~\bibnamefont{Wang}} \bibnamefont{et~al.},
  \bibinfo{journal}{Phys. Rev. Lett.} \textbf{\bibinfo{volume}{101}},
  \bibinfo{eid}{240401} (\bibinfo{year}{2008}).

\bibitem[{\citenamefont{Hofheinz et~al.}(2009)}]{citeulike:4697241}
\bibinfo{author}{\bibfnamefont{M.}~\bibnamefont{Hofheinz}}
  \bibnamefont{et~al.}, \bibinfo{journal}{Nature}
  \textbf{\bibinfo{volume}{459}}, \bibinfo{pages}{546} (\bibinfo{year}{2009}).

\bibitem[{\citenamefont{Sillanp\"{a}\"{a}
  et~al.}(2007)\citenamefont{Sillanp\"{a}\"{a}, Park, and
  Simmonds}}]{2008_Simmonds_TwoQubits}
\bibinfo{author}{\bibfnamefont{M.~A.} \bibnamefont{Sillanp\"{a}\"{a}}},
  \bibinfo{author}{\bibfnamefont{J.~I.} \bibnamefont{Park}}, \bibnamefont{and}
  \bibinfo{author}{\bibfnamefont{R.~W.} \bibnamefont{Simmonds}},
  \bibinfo{journal}{Nature} \textbf{\bibinfo{volume}{449}},
  \bibinfo{pages}{438} (\bibinfo{year}{2007}).

\bibitem[{\citenamefont{Majer et~al.}(2007)}]{majer-2007-449}
\bibinfo{author}{\bibfnamefont{J.}~\bibnamefont{Majer}} \bibnamefont{et~al.},
  \bibinfo{journal}{Nature} \textbf{\bibinfo{volume}{449}},
  \bibinfo{pages}{443} (\bibinfo{year}{2007}).

\bibitem[{\citenamefont{Wallquist et~al.}(2006)\citenamefont{Wallquist,
  Shumeiko, and Wendin}}]{wallquist:224506}
\bibinfo{author}{\bibfnamefont{M.}~\bibnamefont{Wallquist}},
  \bibinfo{author}{\bibfnamefont{V.~S.} \bibnamefont{Shumeiko}},
  \bibnamefont{and} \bibinfo{author}{\bibfnamefont{G.}~\bibnamefont{Wendin}},
  \bibinfo{journal}{Phys. Rev. B} \textbf{\bibinfo{volume}{74}},
  \bibinfo{eid}{224506} (\bibinfo{year}{2006}).

\bibitem[{\citenamefont{Blais et~al.}(2004)\citenamefont{Blais, Huang,
  Wallraff, Girvin, and Schoelkopf}}]{PhysRevA.69.062320}
\bibinfo{author}{\bibfnamefont{A.}~\bibnamefont{Blais}},
  \bibinfo{author}{\bibfnamefont{R.-S.} \bibnamefont{Huang}},
  \bibinfo{author}{\bibfnamefont{A.}~\bibnamefont{Wallraff}},
  \bibinfo{author}{\bibfnamefont{S.~M.} \bibnamefont{Girvin}},
  \bibnamefont{and} \bibinfo{author}{\bibfnamefont{R.~J.}
  \bibnamefont{Schoelkopf}}, \bibinfo{journal}{Phys. Rev. A}
  \textbf{\bibinfo{volume}{69}}, \bibinfo{pages}{062320}
  (\bibinfo{year}{2004}).

\bibitem[{\citenamefont{Blais et~al.}(2007)\citenamefont{Blais, Gambetta,
  Wallraff, Schuster, Girvin, Devoret, and Schoelkopf}}]{blais:032329}
\bibinfo{author}{\bibfnamefont{A.}~\bibnamefont{Blais}},
  \bibinfo{author}{\bibfnamefont{J.}~\bibnamefont{Gambetta}},
  \bibinfo{author}{\bibfnamefont{A.}~\bibnamefont{Wallraff}},
  \bibinfo{author}{\bibfnamefont{D.~I.} \bibnamefont{Schuster}},
  \bibinfo{author}{\bibfnamefont{S.~M.} \bibnamefont{Girvin}},
  \bibinfo{author}{\bibfnamefont{M.~H.} \bibnamefont{Devoret}},
  \bibnamefont{and} \bibinfo{author}{\bibfnamefont{R.~J.}
  \bibnamefont{Schoelkopf}}, \bibinfo{journal}{Phys. Rev. A}
  \textbf{\bibinfo{volume}{75}}, \bibinfo{eid}{032329} (\bibinfo{year}{2007}).

\bibitem[{\citenamefont{Koch et~al.}(2007)\citenamefont{Koch, Yu, Gambetta,
  Houck, Schuster, Majer, Blais, Devoret, Girvin, and
  Schoelkopf}}]{koch-2007-76}
\bibinfo{author}{\bibfnamefont{J.}~\bibnamefont{Koch}},
  \bibinfo{author}{\bibfnamefont{T.~M.} \bibnamefont{Yu}},
  \bibinfo{author}{\bibfnamefont{J.}~\bibnamefont{Gambetta}},
  \bibinfo{author}{\bibfnamefont{A.~A.} \bibnamefont{Houck}},
  \bibinfo{author}{\bibfnamefont{D.~I.} \bibnamefont{Schuster}},
  \bibinfo{author}{\bibfnamefont{J.}~\bibnamefont{Majer}},
  \bibinfo{author}{\bibfnamefont{A.}~\bibnamefont{Blais}},
  \bibinfo{author}{\bibfnamefont{M.~H.} \bibnamefont{Devoret}},
  \bibinfo{author}{\bibfnamefont{S.~M.} \bibnamefont{Girvin}},
  \bibnamefont{and} \bibinfo{author}{\bibfnamefont{R.~J.}
  \bibnamefont{Schoelkopf}}, \bibinfo{journal}{Phys. Rev. A}
  \textbf{\bibinfo{volume}{76}}, \bibinfo{pages}{042319}
  (\bibinfo{year}{2007}).

\end{thebibliography}

\end{document}